\documentclass[twocolumn,showpacs,preprintnumbers,amsmath,amssymb]{revtex4}
\usepackage{graphicx}% Include figure files
\usepackage{dcolumn}% Align table columns on decimal point
\usepackage{bm}% bold math
\usepackage{amsmath}
\usepackage{amssymb}
\usepackage{latexsym}
\usepackage{epsfig}
\usepackage{amsbsy}
\usepackage{array}
\usepackage{amssymb}
\usepackage{setspace}
\newcommand{\upcite}[1]{\textsuperscript{\textsuperscript{\cite{#1}}}}

\begin{document}

\preprint{}

\title{Charge transfer of edge states in zigzag silicene nanoribbons with Stone-Wales defects}% Force line breaks with \\

\author{Wang Rui\footnote{Email: rcwang@cqu.edu.cn.}$^{1,2}$, Wang Shaofeng$^1$, and Wu Xiaozhi$^1$}
\affiliation{$^1$Institute for Structure and Function and
Department of physics, Chongqing University, Chongqing 400044, China.}

\affiliation{$^2$State Key Laboratory of Theoretical Physics,  Institute of Theoretical Physics, Chinese Academy of Science, Beijing 100190, China.}
\date{\today}

\begin{abstract}
Stone-Wales (SW) defects are favorably existed in graphenelike materials with honeycomb lattice structure and potentially employed to change the electronic properties in band engineering. In this paper, we investigate structural and electronic properties of SW defects in bulk silicene and its nanoribbons as a function of their concentration using the methods of periodic boundary conditions with first-principles calculations. We first calculate the formation energy, structural properties, and electronic band structures of SW defects in bulk silicene, with dependence on the concentration of SW defects. Our results show a good agreement with available values from the previous first-principles calculations. The energetics, structural aspects, and electronic properties of SW defects with dependence on defect concentration and location in edge-hydrogenated zigzag silicene nanoribbons are obtained. For all calculated concentrations, the SW defects prefer to locate at the edge due to the lower formation energy. The SW defects at the center of silicene nanoribbons slightly influence on the electronic properties, whereas the SW defects at the edge of silicene nanoribbons split the degenerate edge states and induce a sizable gap, which depends on the concentration of defects. It is worth to find that the SW defects produce a perturbation repulsive potential, which leads the decomposed charge of edge states at the side with defect to transfer to the other side without defect.
\end{abstract}

\pacs{75.70.Ak, 63.22.Np, 61.72.-y, 68.35.Dv}

\keywords{Zigzag silicene nanoribbons; Stone-Wales defects; Electronic properties; First-principles. }%Use showkeys class option if keyword

\maketitle
\baselineskip 13pt

\section{Introduction}

Currently, silicon (Si) based nanostructures, such as two-dimensional (2D) siliene, one-dimensional (1D) silicene nanoribbons, and zero-dimensional (0D) silicene clusters, are emerged as a immense interest material since their properties are similar to but richer than those of graphene\upcite{Cahangirov2009,Jose2012, Lebegue2009}. Similar to graphene, silicene is a monolayer composed of Si atoms in a 2D honeycomb lattice and a linear dispersion near the Fermi level at the $K$ point of the first Brillouin zone (BZ) makes the behavior of charge carriers as massless Dirac fermions. Contrary to graphene, the sublattice inversion symmetry can be broken due to the low-buckled geometry of siliene by applying a perpendicular electric field, leading to a sizable band gap of up to tens of meV\upcite{Ni2012,Drummond2012,Ezawa2012}.  Some $sp^3$ hybridization can be found in silicene in which a large spin-orbit coupling (SOC) are strengthened due to the buckled configuration\upcite{Ezawa2013prl}.
 Recently, the synthesis of silicene on different substrates, such as Ag(111)\upcite{Vogt2012,Feng2012,Jamgotchian2012,Ezawa2013}, (0001)-oriented ZrO2 on Si(111) wafers\upcite{Fleurence2012}, Ir(111)\upcite{Meng2013}, and MoS2 surfaces\upcite{Scalise2014}, etc., has been realized, and facilitates to further investigate the material and its properties. In comparison with graphene, siliene may be better suited for practical electronic applications since it can be more likely to integrate into Si-based electronic devices\upcite{Takeda1994}.  Not only siliene but siliene nanoribbons (SiNRs) have also generated much scientific interest\upcite{Ding2009,Padova2010,Padova2011}.  Analogous to the graphene nanoribbons, SiNRs have two edge types, i.e., zigzag SiNR (ZSiNR) and armchair SiNR (ASiNR).  The first principles calculations indicate that   ASiNRs exhibit metals or semiconductor depending on the ribbon width and ZSiNRs show the antiferromagnetic groundstate \upcite{Ding2009}. Recent investigations have demonstrated that SiNRs, especially ZSiNRs, have rich electronic, transport, magnetic properties and may be applied in silicon-based electronic and spintronic nanodevices potentially\upcite{Le2014,Shakouri2015, Yang2014, Ding2014}.

The topological structural defects in graphene-based materials have been predicted to open an electronic energy band gap that can use in the design of transistors suitably\upcite{Samsonidze2002,Ewels2002,Ma2009,Lusk2008}. Stone-Wales (SW) defect is simplest topological defect which is formed by an in-plane $90^\circ$ rotation of a bond with respect to the midpoint of the bond in 2D honeycomb lattice materials. As a result, a SW defect consists of a pair of pentagons and a pair of heptagons, and can be regarded as a dislocation (pentagon-heptagon ring) dipole\upcite{Ding2007, Wang2011}. The properties of SW defects in graphene had been investigated intensely, and this defect had been observed by a rapid quenching from high temperature or when graphene is under irradiation experimentally \upcite{Meyer2008}. In silicene, some studies to understand the electronic properties, formation, stability relative to perfect silicene, and reactivity of SW defects had been presented\upcite{Sahin2013}. In comparison with the formation energy of SW defects in graphene (4.66$\sim$5.82 eV)\upcite{Ma2009, Shirodkar2012}, the formation energy in silicene are found to be 1.64$\sim$2.09 eV\upcite{Sahin2013,Manjanath2013,Gao2013}, much smaller than that in graphene. It means that a SW defect can be more easily formed in siliene than in graphene. However, there have been only a few investigations on the properties of SW defects in silicene\upcite{Sahin2013,Manjanath2013,Gao2013} and its nanoribbons\upcite{Dong2015}. In this paper, we use first-principles calculations with the density functional theory (DFT) to investigate the structural and electronic properties of SW defects with dependence on the concentration and location in edge-hydrogenated ZSiNRs.  The results show that the SW defects prefer to locate at the edge due to the lower formation energy for all calculated concentrations. The SW defects at the center of ZSiNRs slightly influence on the electronic properties, however, the SW defects at the edge of ZSiNRs induce a sizable gap in the band dispersions and edge localized states are present. It is interesting to find that the SW defects at the edge results in the decomposed charge of the edge states transferring to the other side opposite of the side with defect. Our results are potentially used to design the electronic devices based on silicene in future band engineering.

Our paper is organized as follows. Sec. \ref{Com} briefly describes the computational methodology. In Sec. \ref{SWbul}, the structural and electronic properties of SW defects with dependence on concentration of defects in bulk silicene are studied. Sec. \ref{SWNR} presents the structural and electronic properties in SW-defected ZSiNRs and their dependence on the location and concentration of defects. Finally, we conclude in Sec. \ref{con}.

\section{Computational methodology}\label{Com}
First principles calculations are carried out within density functional theroy (DFT) formalism\upcite{Hohenberg,Kohn} as implemented in the Vienna \textit{ab initio} simulation package (VASP) \upcite{Kresse2,Kressecom}.  The Kohn-Sham (KS) equations are solved self-consistently within the projector augmented wave (PAW) method\upcite{Blochl,Kresse4} and the plane-wave basis set with a cutoff energy of 500 eV. The exchange-correlation functionals are described by local density approximation (LDA)\upcite{Ceperley1980} and the Perdew-Burke-Ernzerhof (PBE) generalized gradient approximation (GGA)\upcite{Perdew1,Perdew2} for calculations of bulk silicene with and without SW defects, and only LDA for calculations of silicene nanoribbons. We included a vacuum region of 18 {\AA} to avoid the interaction between the periodic layers and ribbons, respectively. In order to determine the equilibrium geometry of bulk silicene and nanoribbons with defects, all the atomic coordinates and supercell configurations are relaxed by the conjugate gradient algorithm until an energy convergence of $10^{-6}$eV and a force convergence of  0.01eV/{\AA}. The sampling of the Brillouin zone (BZ) has been done for the two-dimensional (2D) supercell with the equivalent of $25\times25\times1$ Monkhorst-Pack\upcite{Monkhorst} special k-point scheme for a bulk silicene primitive cell, and the one-dimensional (1D) supercell with the equivalent of $25\times1\times1$ Monkhorst-Pack special k-point scheme for a silicene nanoribbon primitive cell. For the calculations of electronic density of state (DOS), the tetrahedron integration method is employed with a smearing with of 0.05eV.

\section{Stone-Wales defects in bulk silicene}\label{SWbul}
2D honeycomb structures of Si (silicene) have been found stable in a slightly buckled geometry in DFT calculations \upcite{Cahangirov2009} and experiments \upcite{Vogt2012,Feng2012,Jamgotchian2012,Ezawa2013,Fleurence2012,Meng2013,Scalise2014}.  Due to silicene's hexagonal structure, originated from the $sp^2$ hybridization, the grain boundaries are expected to be formed of pentagon-heptagon pairs (i. e., a dislocation dipole), known as stone-Wales (SW) defects. As shown in FIG.\ref{psw}, a SW defect is formed by a $90^\circ$ rotation of a Si dimer, therefore transforming a set of four hexagons into a pentagon and a heptagon pair.

\vspace{0cm}
\makeatletter
\def\@captype{figure}
\centerline{\scalebox{0.8}[0.8]{\includegraphics{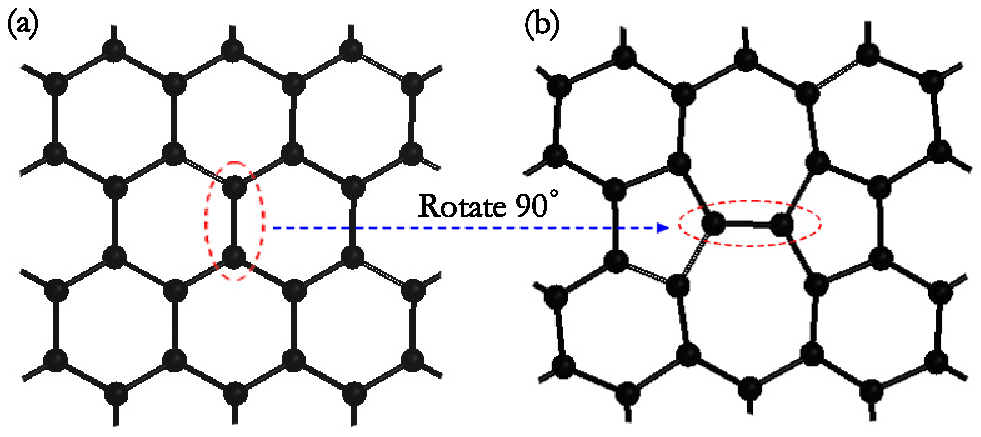}}}
\makeatother
\caption{\footnotesize{(Color online) The relaxed structures of (a) perfect and (b) SW-defected silicene. A SW defect can be created by the rotation of a silicon dimer by $90^\circ$ around the center of the Si-Si bond (the two si atoms in the dashed line circle). }}
\label{psw}

The formation energy, $E_{f}^{\mathrm{SW}}$, of a SW defect in silicene is given by
\begin{equation}
E_{f}^{\mathrm{SW}}=E_{\mathrm{silicene}}^{\mathrm{SW}}-E_{\mathrm{silicene}},
\end{equation}
where, $E_{\mathrm{silicene}}^{\mathrm{SW}}$ and $E_{\mathrm{silicene}}$ are the total energy of silicene with and without SW defect, respectively. We employ the supercell approach and periodic boundary condition to simulate the isolated SW defect in bulk silicene. The interaction between two SW defects is expected to be mediated by strain and long ranged because a SW defect involves a dipole of dislocation\upcite{Wang2011,Shirodkar2012}. In order to evaluate a reasonable size of supercell in which the interaction between two SW defects would be neglected, we choose $4\times 4$, $5\times 5$, $6\times 6$, and $7\times 7$ supercells to simulate the SW defects.

TABLE \ref{FE} lists the calculated results with dependence on concentration of defects in both the cases of LDA and GGA formalisms, compared with the previous  available values\upcite{Sahin2013,Manjanath2013,Gao2013}.  Between the SW defects in $4\times4$ and $7\times7$ supercells, the differences of formation energy $E_{f}^{\mathrm{SW}}$ both for the LDA and GGA are larger than $\sim$ 0.1eV, however those between $5\times5$ and $7\times7$ supercells reduce to $\sim$ 0.03eV. Therefore, the interaction between two SW defects in $5\times5$ supercell is negligible. It corresponds to the system with a defect concentration of $2.7\times10^{13}$ $\mathrm{cm}^{-2}$. In the GGA formalism, the formation energy $E_{f}^{\mathrm{SW}}$  of a SW defect in a $6\times 6$ supercell is found to be 1.63 eV, which is a good agreement with the previous value 1.64 eV obtained by Sahin, et. al.\upcite{Sahin2013}. However, $E_{f}^{\mathrm{SW}}$ in a $7\times 7$ supercell is calculated to be 1.62 eV, slightly lower than the result 1.82 eV calculated by Manjanth, et. al.\upcite{Manjanath2013}. In comparison with the formation energy of SW defects in graphene (4.66$\sim$5.82 eV)\upcite{Ma2009, Shirodkar2012}, a SW defect is much easier formed in silicene due to the buckled lattice geometry.  The length of the center Si dimer ($L_{\mathrm{dimer}}$) is a very important structural parameter, and also listed in TABLE \ref{FE}.  The values of $L_{\mathrm{dimer}}$ both for the LDA and GGA cases seem to be no difference when the size of supercell is larger than $4\times 4$. Hence, the $5\times 5$ or larger supercells can simulate the formation of the local SW defect in silicene appropriately. It is worth to note that the Si-Si bond length compresses both for LDA (from 2.25 to 2.16 \AA) and GGA (from 2.28 to 2.18 \AA) after the formation of SW defect, i.e., the Si-Si bond becomes stronger than that in perfect silicene through  $90^\circ$ rotation of a dimer. For all supercell sizes, the LDA and GGA results exhibit slight differences. The LDA results show slightly higher $E_{f}^{\mathrm{SW}}$ and lower $L_{\mathrm{dimer}}$. In the present work, we are mostly concerned with the electronic structures, thus we report the LDA results on the electronic properties in the following.

\begin{table}
\footnotesize
\caption{\footnotesize{ The calculated results for SW-defected silicene with respect to supercell size using both the LDA and GGA: formation energy $E_{f}^{\mathrm{SW}}$ (eV), and the length of the center Si dimer $L_{\mathrm{dimer}}$ (\AA).  The available results from the previous DFT calculations are also shown for comparison. The data in the parenthesis are the Si-Si bond length in perfect silicene.}}
\begin{tabular}{cccccc}
  \hline
   & & 4$\times$4 & 5$\times$5 & 6$\times$6 & 7$\times$7 \\
  \hline
   $E_{f}^{\mathrm{SW}}$  & LDA & 1.60 & 1.76 & 1.72 & 1.73 \\
            & GGA & 1.51 & 1.65, 2.09\upcite{Gao2013} & 1.63, 1.64\upcite{Sahin2013} & 1.62, 1.82\upcite{Manjanath2013} \\
  \hline
   $L_{\mathrm{dimer}}$ & LDA (2.25) & 2.18 & 2.16 &2.16 & 2.16 \\
                        & GGA (2.28) & 2.21 & 2.18 &2.18, 2.19\upcite{Sahin2013} & 2.18, 2.20\upcite{Manjanath2013}\\

  \hline
\end{tabular}
\label{FE}
\end{table}

\vspace{0.7cm}
\makeatletter
\def\@captype{figure}
\centerline{\scalebox{0.23}[0.28]{\includegraphics{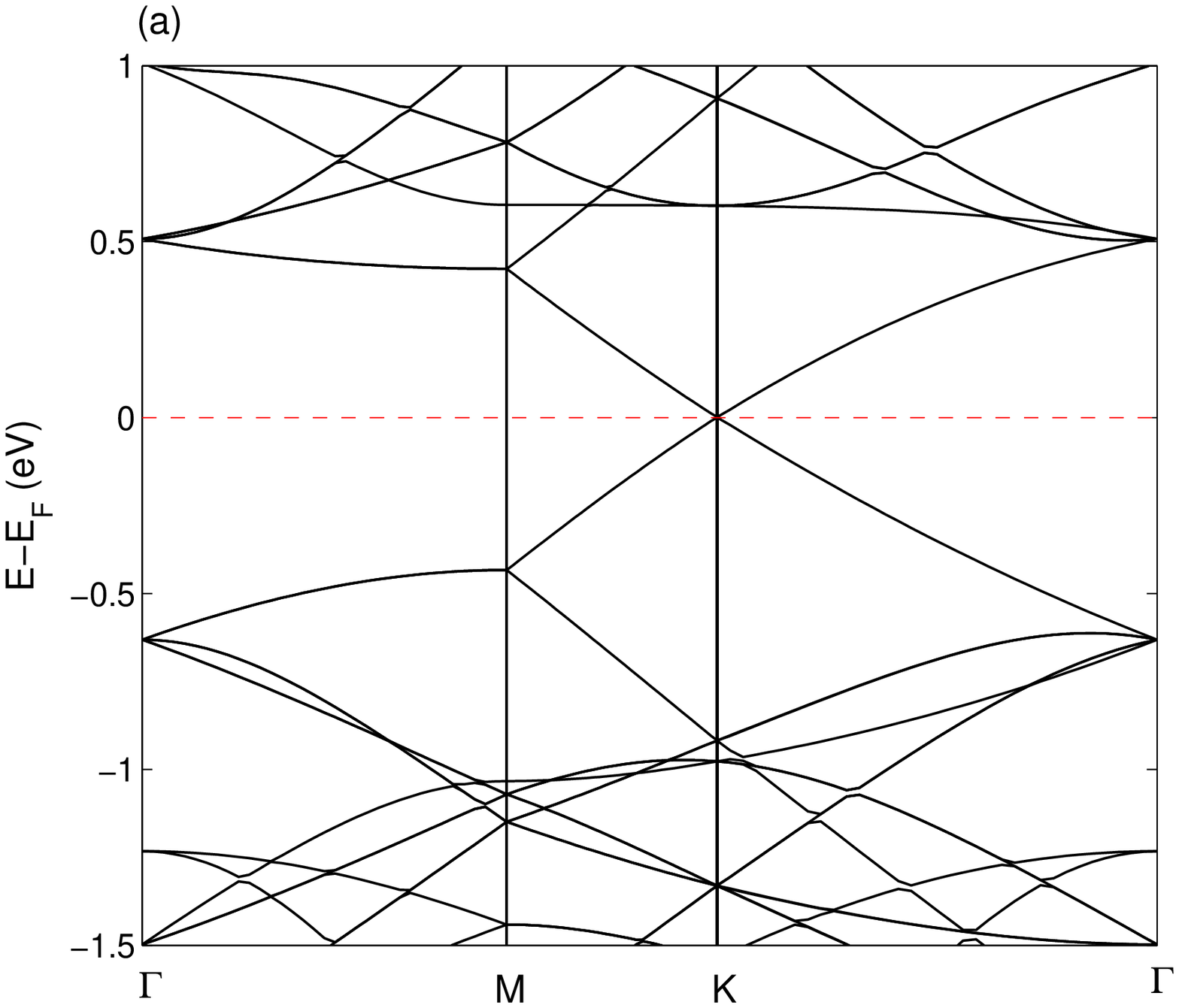}}
\scalebox{0.23}[0.28]{\includegraphics{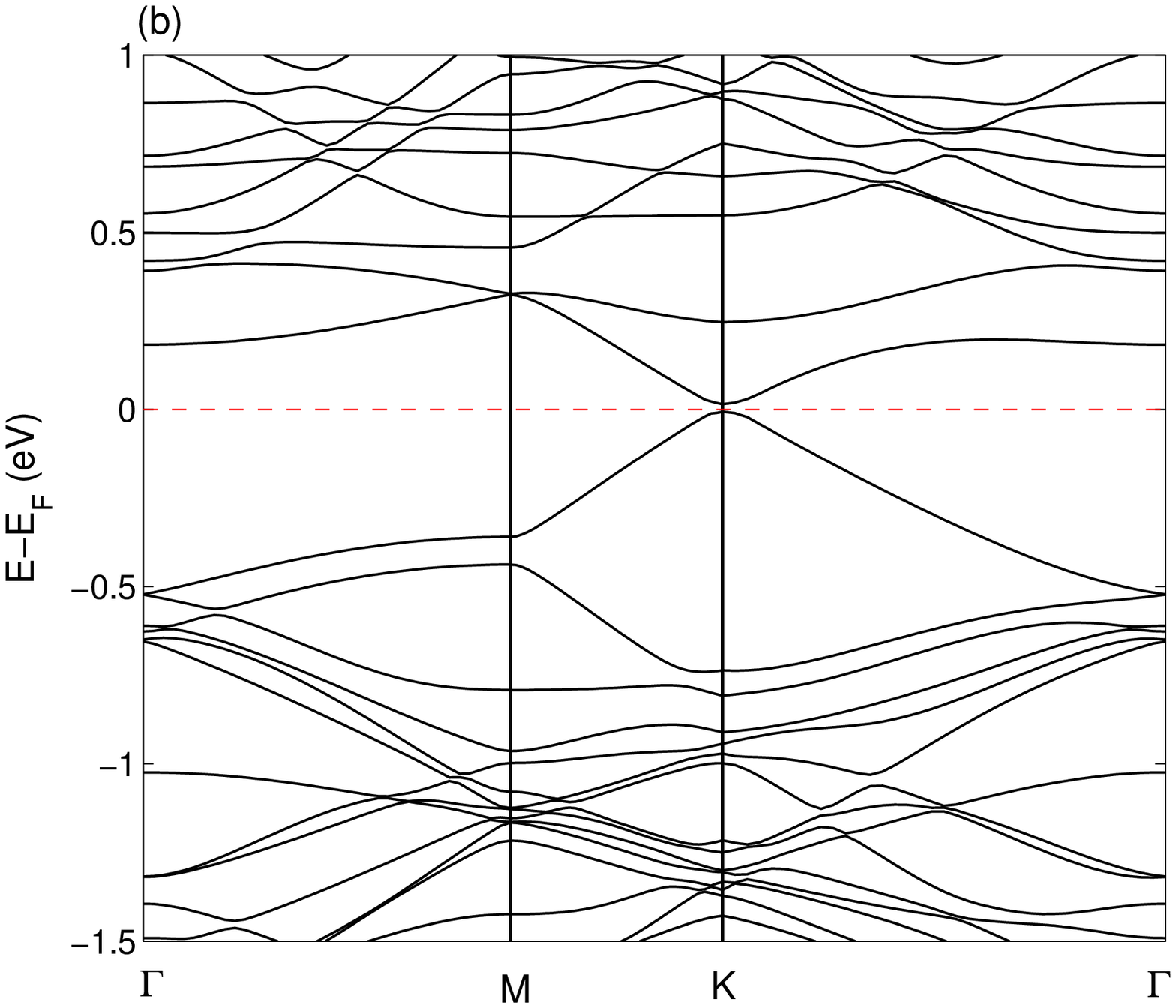}}}
\makeatother
\caption{\footnotesize{(Color online) The electronic band structures for a $5\times5$ supercell of (a) perfect and (b) SW-defected silicene.} }
\label{band55}

\vspace{0.7cm}
\makeatletter
\def\@captype{figure}
\centerline{\scalebox{0.23}[0.28]{\includegraphics{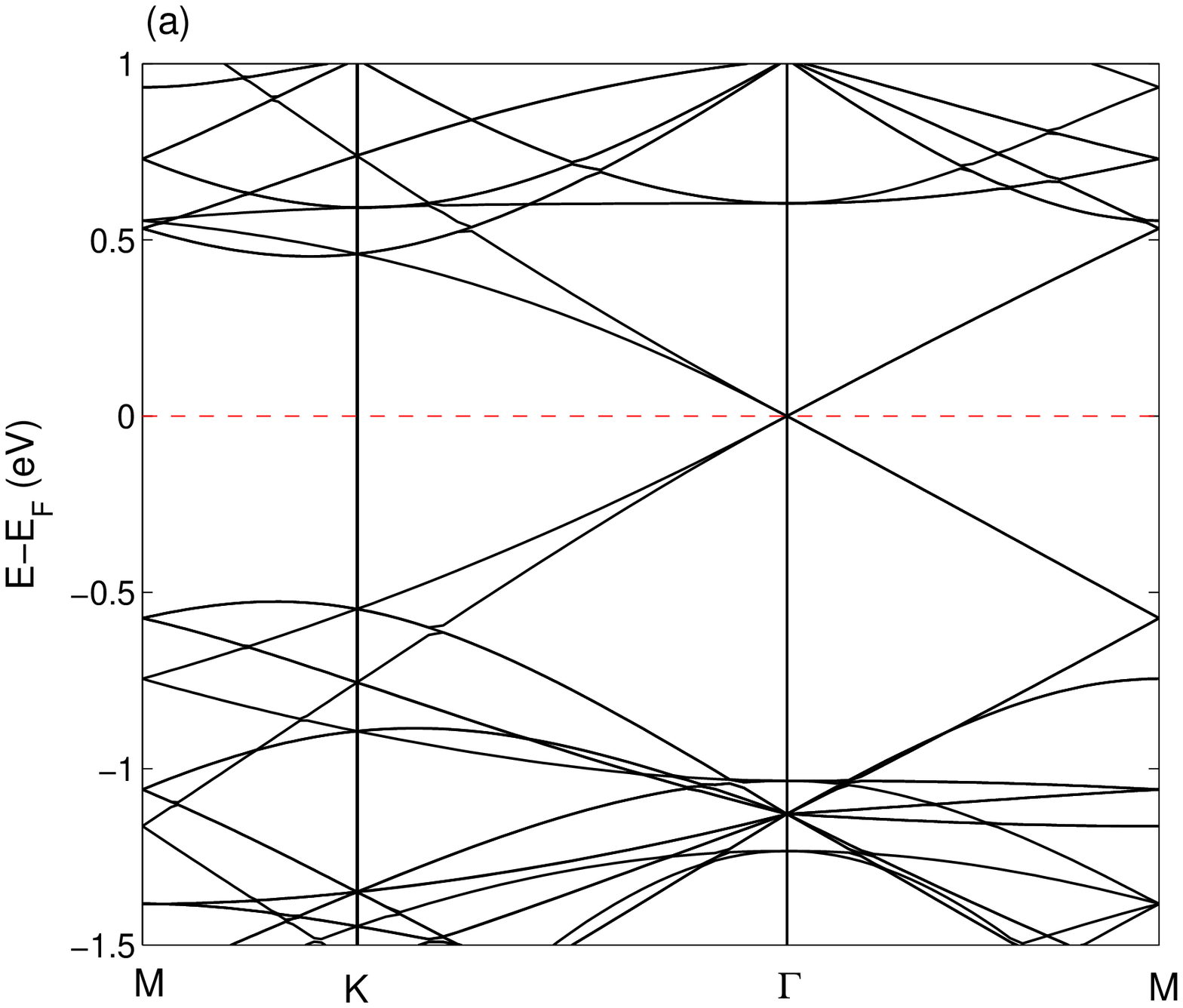}}
\scalebox{0.23}[0.28]{\includegraphics{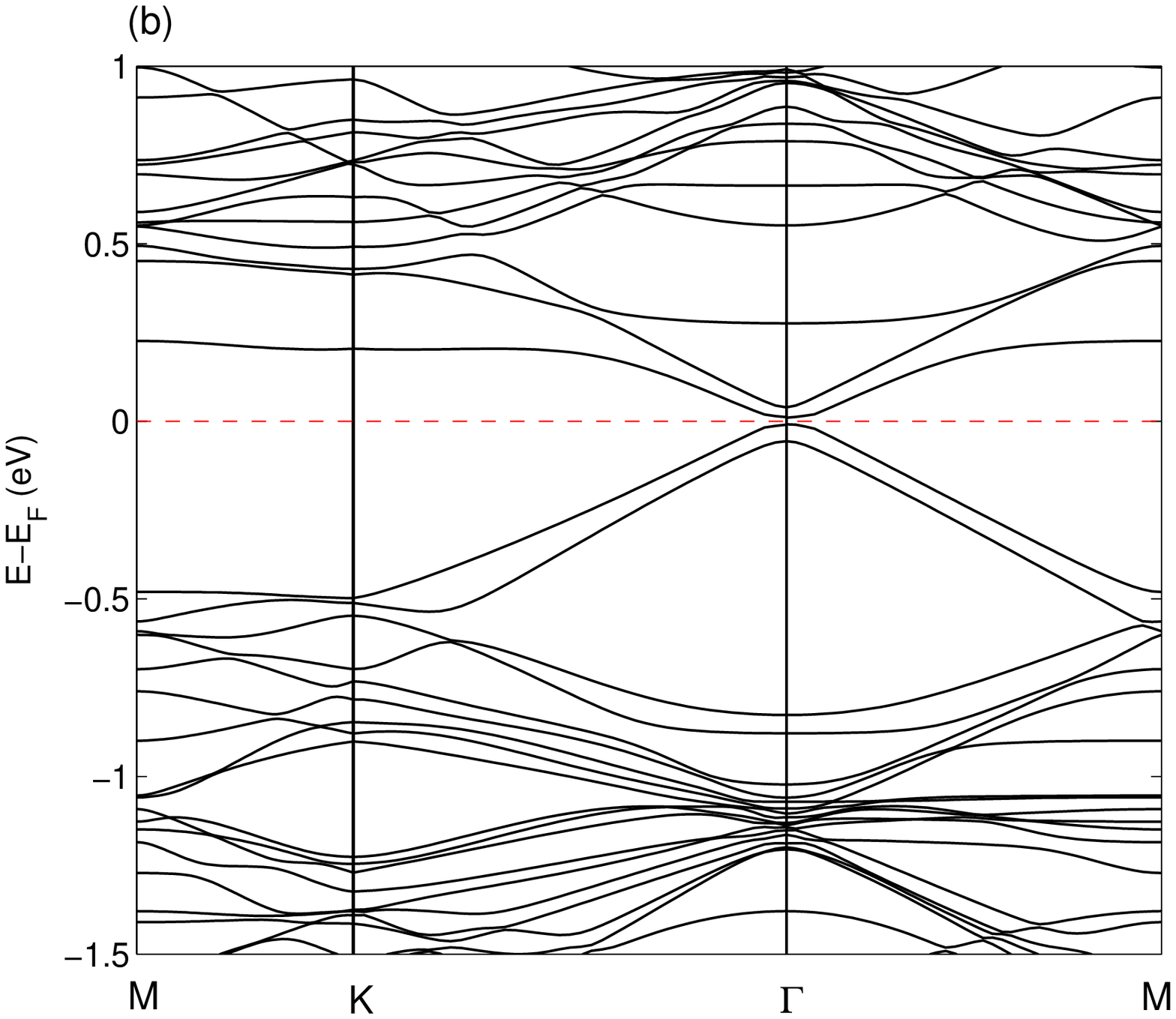}}}
\makeatother
\caption{\footnotesize{(Color online) The electronic band structures for a $6\times6$ supercell of (a) perfect and (b) SW-defected silicene. It is worth indicating that the $K$ and $K'$ of the primitive cell of silicene are folded onto the $\Gamma$ point.}  }
\label{band66}

Next, we investigate the effect of SW defect on the electronic structure of bulk silicene.  In FIGs. \ref{band55} and \ref{band66}, the electronic band dispersions of perfect and SW-defected silicene for $5\times5$ and $6\times6$ supercells are presented, respectively. It is worth noting that the $K$ and $K'$ symmetry points of the primitive cell of silicene are folded onto the $\Gamma$ point for a $6\times6$ supercell, as well as the band edge still remains on the $K$ symmetry point for $5\times5$ supercell. The perfect silicene is a semimetal with a band profile having linear dispersion near the Fermi level, which attributes to its electrons a massless Dirac fermion behavior as shown in FIGs. \ref{band55}(a) and \ref{band66}(a). The profile of band structures for $6\times6$ supercell agrees well with the previous DFT calculations\upcite{Sahin2013}.  It is clear to see that the lattice symmetry of silicene is broken after SW defects are present.  The formation of the SW defect breaks the sixfold symmetry of the silicene lattice and there is a band gap opening at the crossing point\upcite{Sahin2013}. The band gap opening for SW-defected silicene is calculated to be 0.03 eV in a $5\times5$ supercell and 0.02 eV in a $6\times6$ supercell. In fact, the band gap is form the periodic potential of the array of SW defects.

\section{Stone-Wales defects in zigzag silicene nanoribbons}\label{SWNR}

Silicene nanoribbon (SiNR) is a quasi-one dimensional (1D) derivatives of silicene. In this section, we consider the zigzag SiNR (ZSiNR) of width $n=16$ (ZSiNR-16) and $n$ is the counting number of Si atoms forming a armchair chain perpendicular to the nanoribbon axis. Accordingly, there are 16 Si atoms in the primitive unit cell of a zigzag nanoribbon as shown in FIG. \ref{cellribbon}(a), where the edges are saturated by hydrogens.  We first carry out the structural relaxation of the ZSiNR-16 with the edges saturated by hydrogens. Similar to graphene nanoribbon, the $\pi$ orbital of the edge silicon atom (Si-$p_{z}$) in ZSiNR-16 may represent edge states which are strongly localized at the edge silicon atoms \upcite{Lee2009}. Therefore, three kinds of spin-polarized calculation are also performed: nonspin polarization (NM), ferromagnetic (FM) spin coupling for electrons, and antiferromagnetic (AFM) ordering with antiparallel spin orientation between two edges.  The spin magnetic moment is zero for AFM, and 0.58$\mu_{B}$/unit cell for FM.  The band structures of NM, FM, and AFM  are shown in FIG. \ref{cellribbon}(b).  It is clear to see that the valence ($\pi$-top ) and conductance ($\pi^*$-bottom) bands of NM state are flat and degenerate near the $X$ point. The flat region of these $\pi$-top and $\pi^*$-bottom bands is due to that the Dirac point of silicene which folds into $k=2\pi/3$ in the BZ of ZSiNR.  The results agree well with the previous calculations in graphene\upcite{Lee2009} and silicene\upcite{Ding2009} nanoribbons. One can see that the FM state is metallic since either one of spin-up or -down band tends to be occupied to cause the system to be ferromagnetic\upcite{Lee2009}. The $\pi$-top and $\pi^*$-bottom bands of AFM state move down and up, respectively, resulting in a semiconductor. However, the energy differences among NM, FM, and AFM spin configurations are only several meV/unit cell ($E_{\mathrm{FM}}=-9.3$ meV/unit cell, $E_{\mathrm{AFM}}=-9.6$ meV/unit cell relative to $E_{\mathrm{NM}}$). Therefore, the AFM semiconductor is the ground state for ZSiNR at T=0 K. The results are in good agreements with previous DFT calculations by Ding and Ni\upcite{Ding2009}, but the values of $E_{\mathrm{FM}}$ and $E_{\mathrm{AFM}}$ for ZSiNR-16 in this calculation are only about half of those for ZSiNR-12\upcite{Ding2009}. However, the thermal fluctuations will break the FM and AFM spin configurations and induce the NM ground state in the room temperature.  In the following discussion, we only investigate the properties of SW defect in the ZSiNR-16 with NM spin configuration.

In FIG. \ref{cellribbon}(c), the projected density of states (pDOS) of the Si1, Si2, and Si3 atoms which are denoted in FIG. \ref{cellribbon}(a) for the NM state are listed respectively. The $p_z$ orbital of the edge silicon atom (Si1) shows the largest contribution at the Fermi level and Si3-$p_z$ also exhibits a peak but it is much lower than that of Si1 atom, whereas Si2-$p_z$ does not contribute at all. The bar graph of DOS of $p_z$ orbital at Fermi level as a function of the position of silicon atom are shown in FIG . \ref{cellribbon}(d). The result indicates that the states in the flat band region representing edge states only come from the odd silicon positions and are mainly localized at the edge Si atoms.
\vspace{0.7cm}
\makeatletter
\def\@captype{figure}
\centerline{\scalebox{0.67}[0.67]{\includegraphics{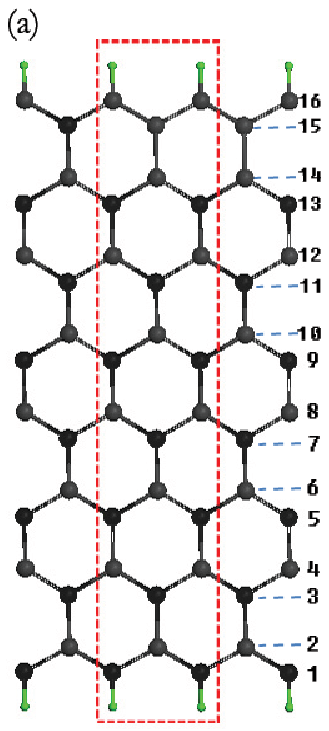}}
\scalebox{0.33}[0.33]{\includegraphics{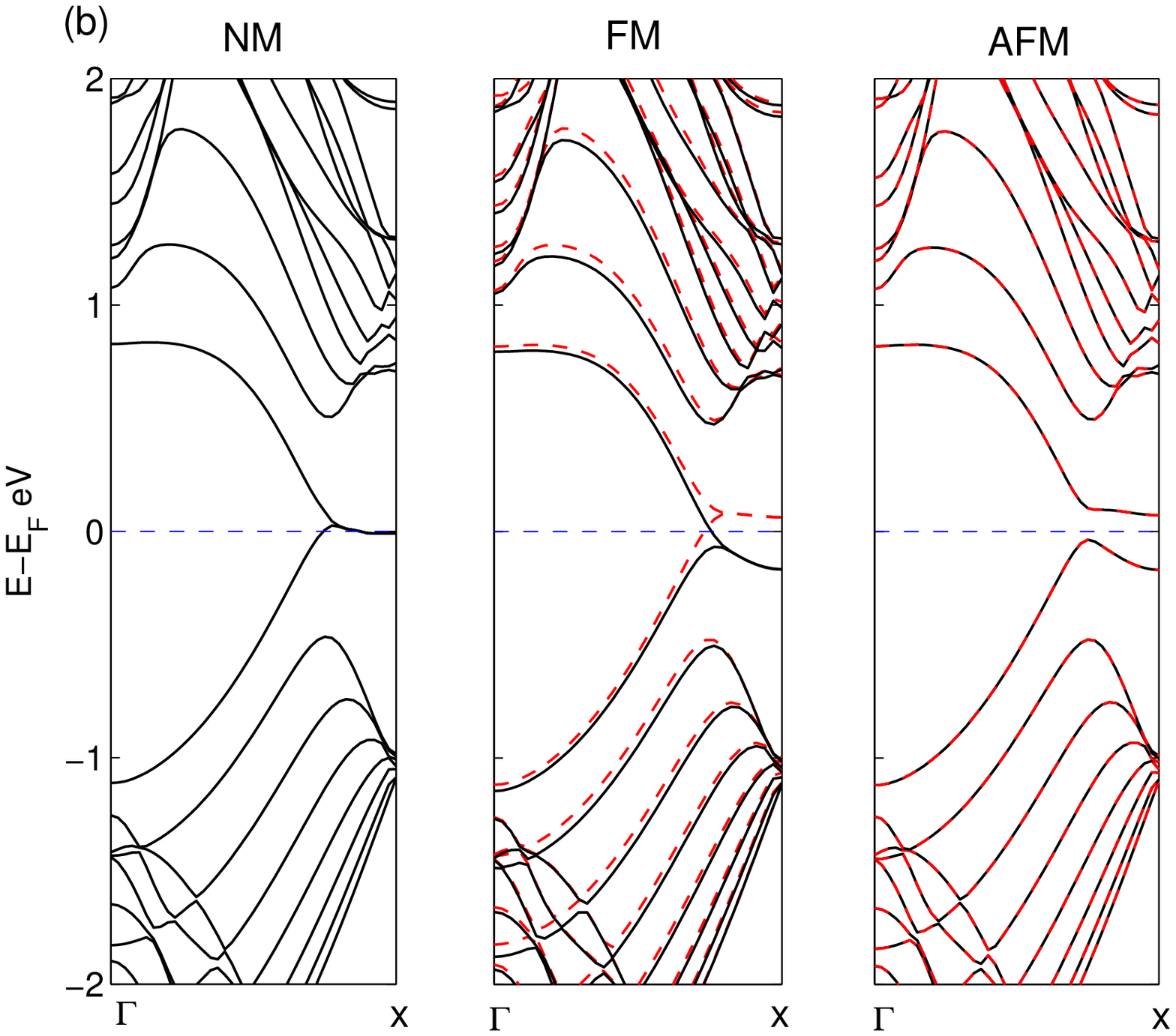}}}
\centerline{\scalebox{0.29}[0.29]{\includegraphics{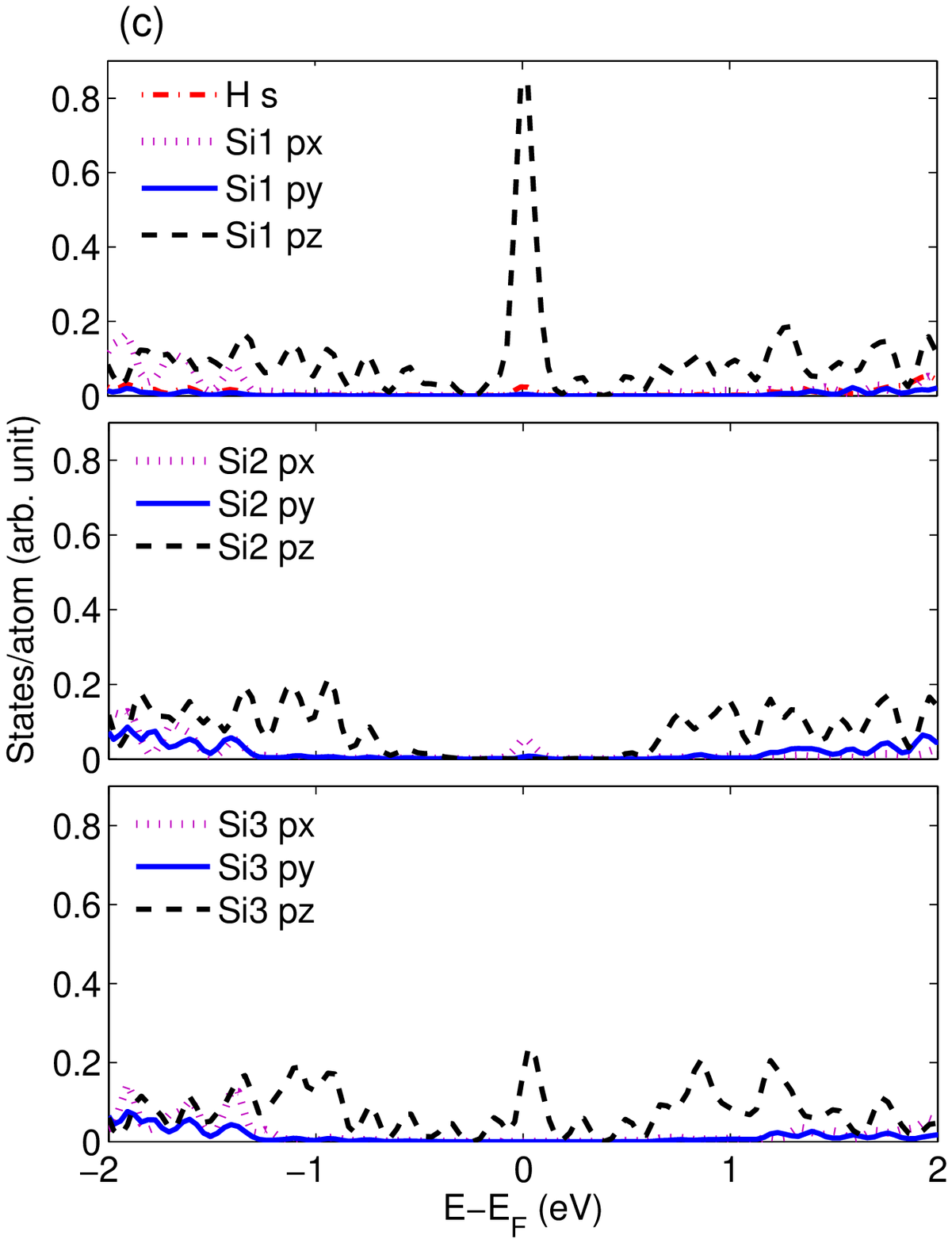}}
\scalebox{0.24}[0.30]{\includegraphics{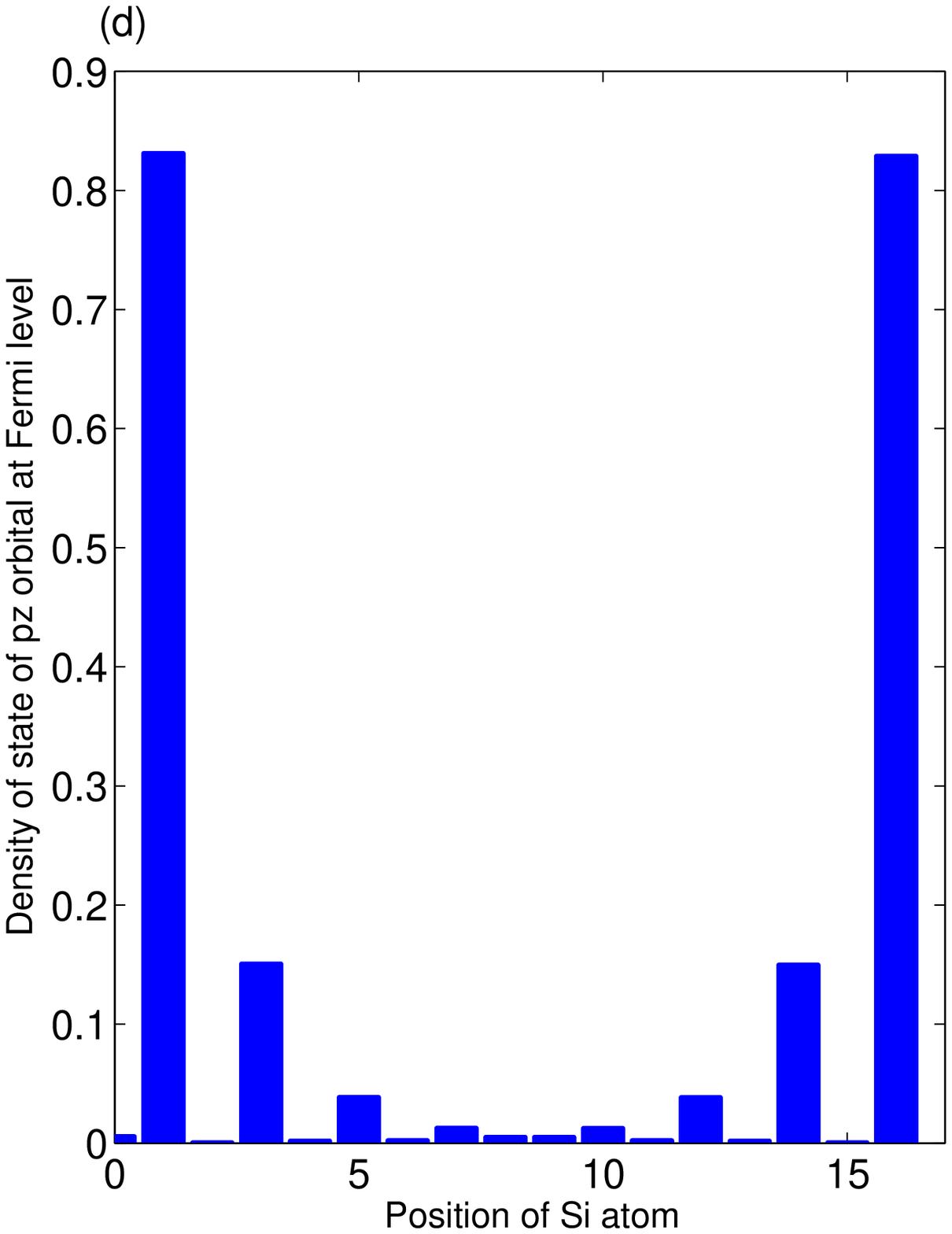}}}
\makeatother
\caption{\footnotesize{(Color online) (a) The relaxed structure of hydrogen saturated ZSiNR-16. The primitive unit cell of ZSiNR-16 is indicated by dashed line box. (b) The band structures for NM, FM, and AFM  spin configuration are shown. (c) The projected density of states for Si1, Si2, and Si3 atoms for the NM state. (d) The DOS value of $p_z$ orbital at Fermi level as a function of the position of silicon atom.}  }
\label{cellribbon}

Next, we investigate structural and electronic properties of ZSiNR-16 with different concentrations of SW defects in the NM spin configuration. We choose $4\times1$, $5\times1$,  and $6\times1$ supercells based on the primitive cell of ZSiNR-16 as shown in FIG. \ref{cellribbon} (a) to investigate size effect on the properties of SW defect. The unit cell consists of 64, 80, and 96 Si atoms for $4\times1$, $5\times1$, and $6\times1$ supercells, respectively, with additional H atoms attached to chemically modify the edges. There are two different typical defect positions in the supercell to be considered: at the center and at the edge. For instance, the relaxed structure of perfect, SW defect at the center, and SW defect at the edge in $6\times1$ supercell are shown in FIG. \ref{ribbon-sw} . In comparison with the atomic structures of ZSiNR-16 with and without SW defect, a compressive stress field arises from the ribbon with defect. The SW defect is a dipole consisted of neighboring dislocation (pentagon-heptagon ring) and anti-dislocation\upcite{Ding2007, Wang2011}. Therefore, the compressive stress field results from the attraction between dislocation and anti-dislocation.

\begin{table}
\caption{\footnotesize{ Calculated structural and electronic properties for SW-defected ZSiNR-16: formation energy $E_{f}^{\mathrm{SW}}$ (eV),  Si dimer $L_{\mathrm{dimer}}$ (\AA), and band gap $E_{\mathrm{gap}}$ (eV) of SW defect at the center and edge in $4\times1$, $5\times1$, and $6\times1$ supercell, respectively. }}
\begin{tabular}{ccccc}
  \hline
  \ \ & \ \ \ & \ \ \ $4\times1$  \ \ \ \ \ \ &  \ \ \ \ \ \ $5\times1$  \ \ \ \ \ \ &  \ \ \ \ \ \ $6\times1$  \ \ \ \ \ \ \\
  \hline
  $E_{f}^{\mathrm{SW}}$ & center  & 1.97 & 1.85 & 1.76 \\
  \ \                   & edge    & 1.89 & 1.66 & 1.59 \\
  \hline
  $L_{\mathrm{dimer}}$ &  center  & 2.156 & 2.158 & 2.163\\
  \ \                  &  edge    & 2.159 & 2.165 & 2.170\\
  \hline
  $E_{\mathrm{gap}}$   &  center  & 0.17 & 0 & 0\\
  \ \                  &  edge    & 0.33 & 0.23 & 0.15\\
  \hline
\end{tabular}
\label{ribbonswdata}
\end{table}

\vspace{0.2cm}
\makeatletter
\def\@captype{figure}
\centerline{
\scalebox{0.35}[0.35]{\includegraphics{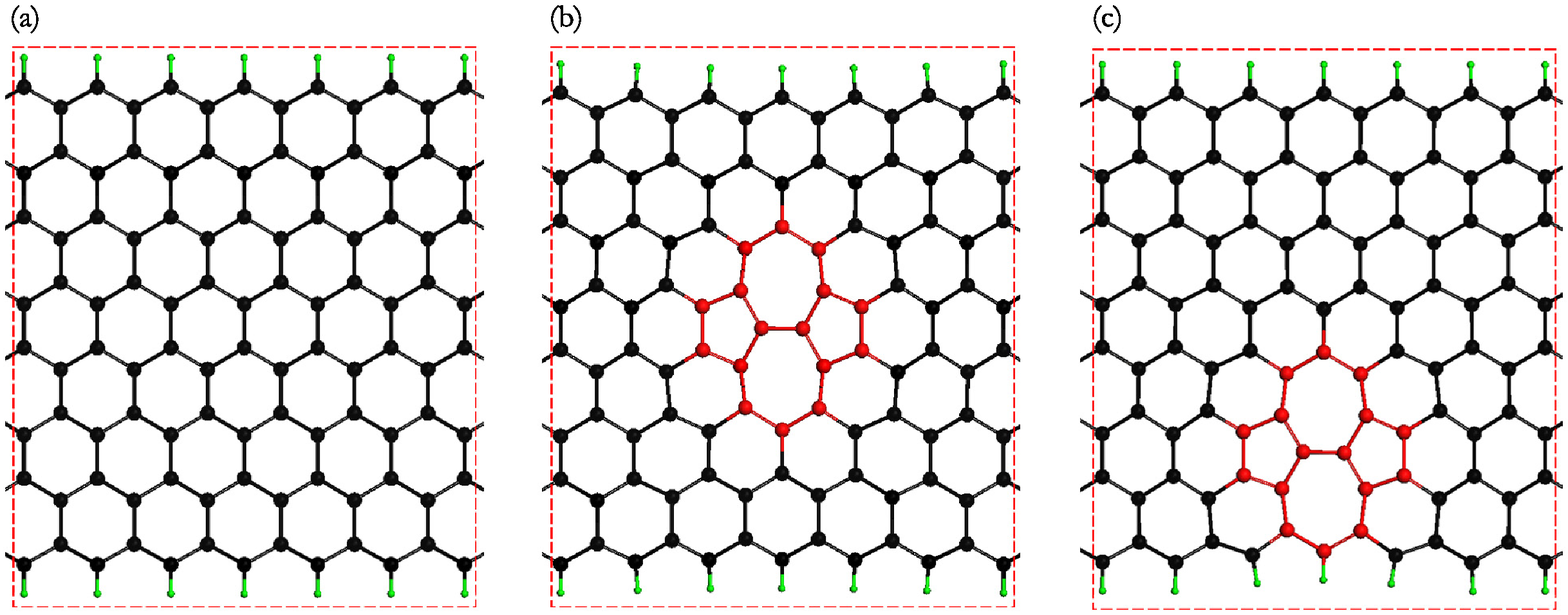}}}
\makeatother
\caption{\footnotesize{(Color online) The relaxed structures of (a) perfect ZSiNR-16, (b) SW defect at the center, and (c) SW defect at the edge for $6\times1$ supercell. The region in the dashed box describes the basic cell (i. e., $6\times1$ supercell) in the first-principles calculation. The edge silicon atoms are saturated by hydrogen atoms. }  }
\label{ribbon-sw}

The calculated structural and electronic properties are listed in TABLE \ref{ribbonswdata}. The length of Si dimer $L_{\mathrm{dimer}}$ at the center is slightly lower than that at the edge. With increasing supercell size, the length of Si dimer $L_{\mathrm{dimer}}$ increase and this means the compressive stress field weakening. It is interesting to find that the formation energies $E_{f}^{\mathrm{SW}}$ of SW defect at the edge are always lower than that at the center in ZSiNR-16, and the results indicate that it is easier for SW defect to be formed at the edge than the center. The calculated results are in a agreement with previous first principles calculations on SW defect in the silicene nanosheet\upcite{Manjanath2013}. The formation energies of SW defect in ZSiNR-16 decrease with increasing the size of supercell, since there is lower defect density in large supercell. In addition, the formation energy of SW defect at the edge in ZSiNR-16 with $6\times1$ supercell is 1.59 eV, which is lower than that in bulk silicene with $6\times6$ supercell 1.72 eV. Therefore, we can more easily obtain the SW defect in silicene nanoribbon that bulk silicene.
\vspace{0.1cm}
\makeatletter
\def\@captype{figure}
\centerline{
\scalebox{0.4}[0.4]{\includegraphics{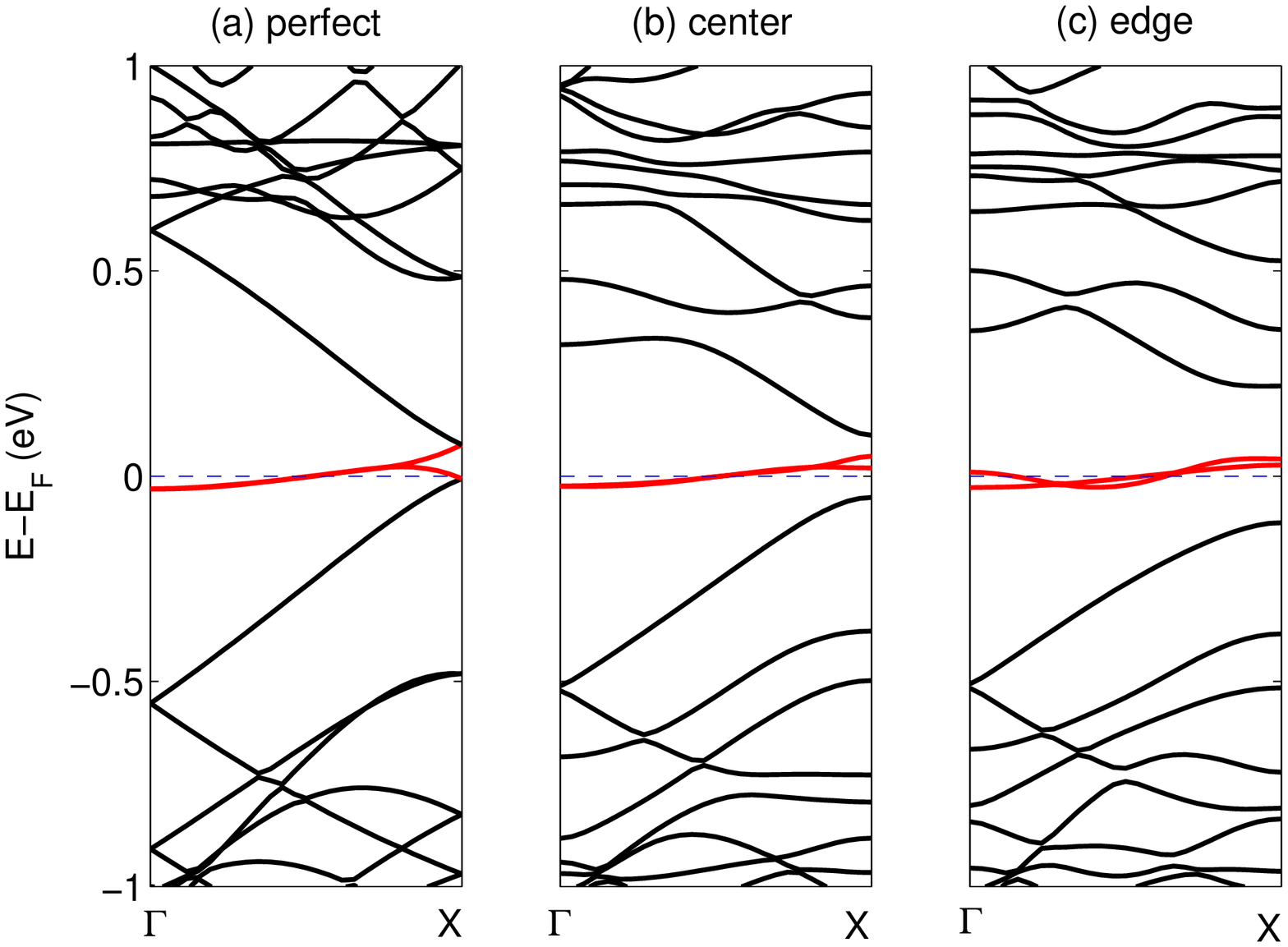}}}
\centerline{\scalebox{0.48}[0.48]{\includegraphics{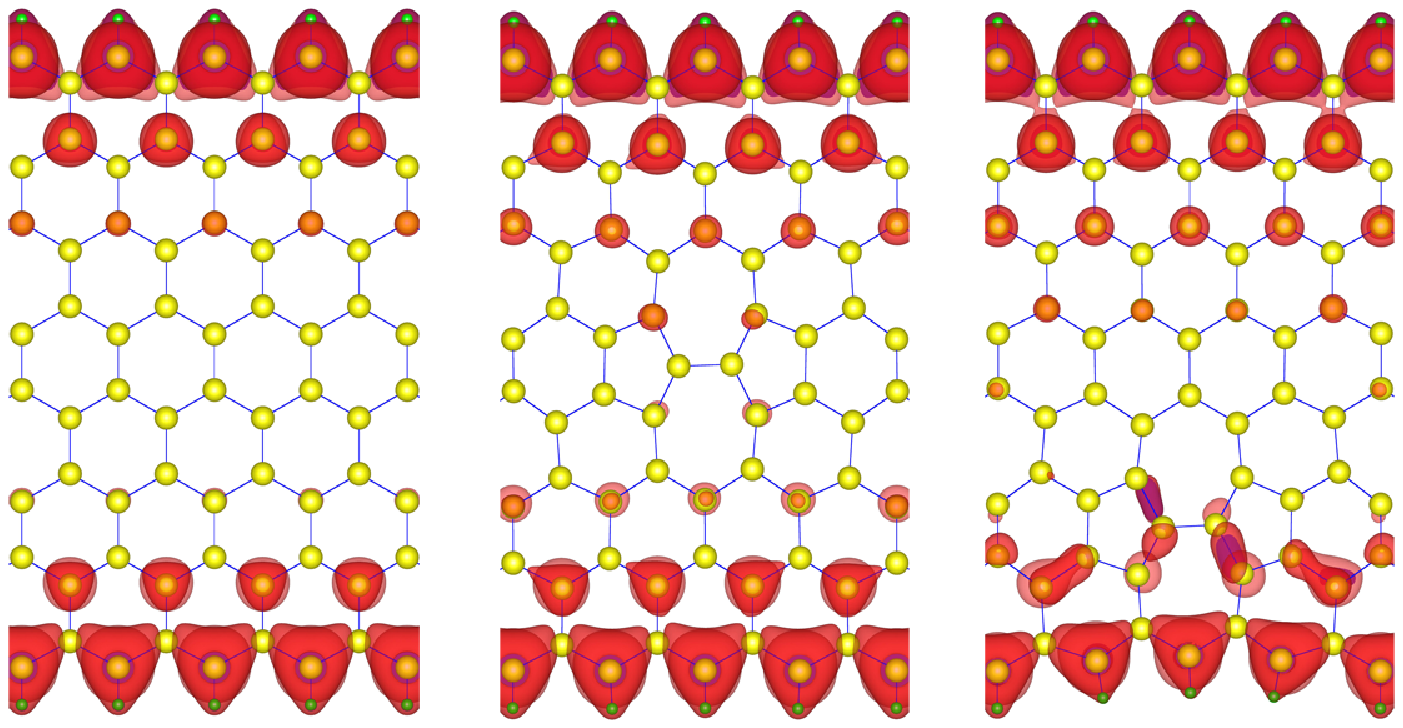}}}
\makeatother
\caption{\footnotesize{(Color online) The band structures of ZSiNR-16 for $4\times1$ supercell of (a) perfect, (b) SW defect at the center, and (c) SW defect at the edge. The solid red lines indicate the edge state. Band decomposed charge densities of the edge states are also shown at the bottom of corresponding band structures (the isosurface is set to $0.55\times 10^{-3} $ e/$\mathrm{\AA} ^3$).}  }
\label{band16-4}

\vspace{0.1cm}
\makeatletter
\def\@captype{figure}
\centerline{
\scalebox{0.4}[0.4]{\includegraphics{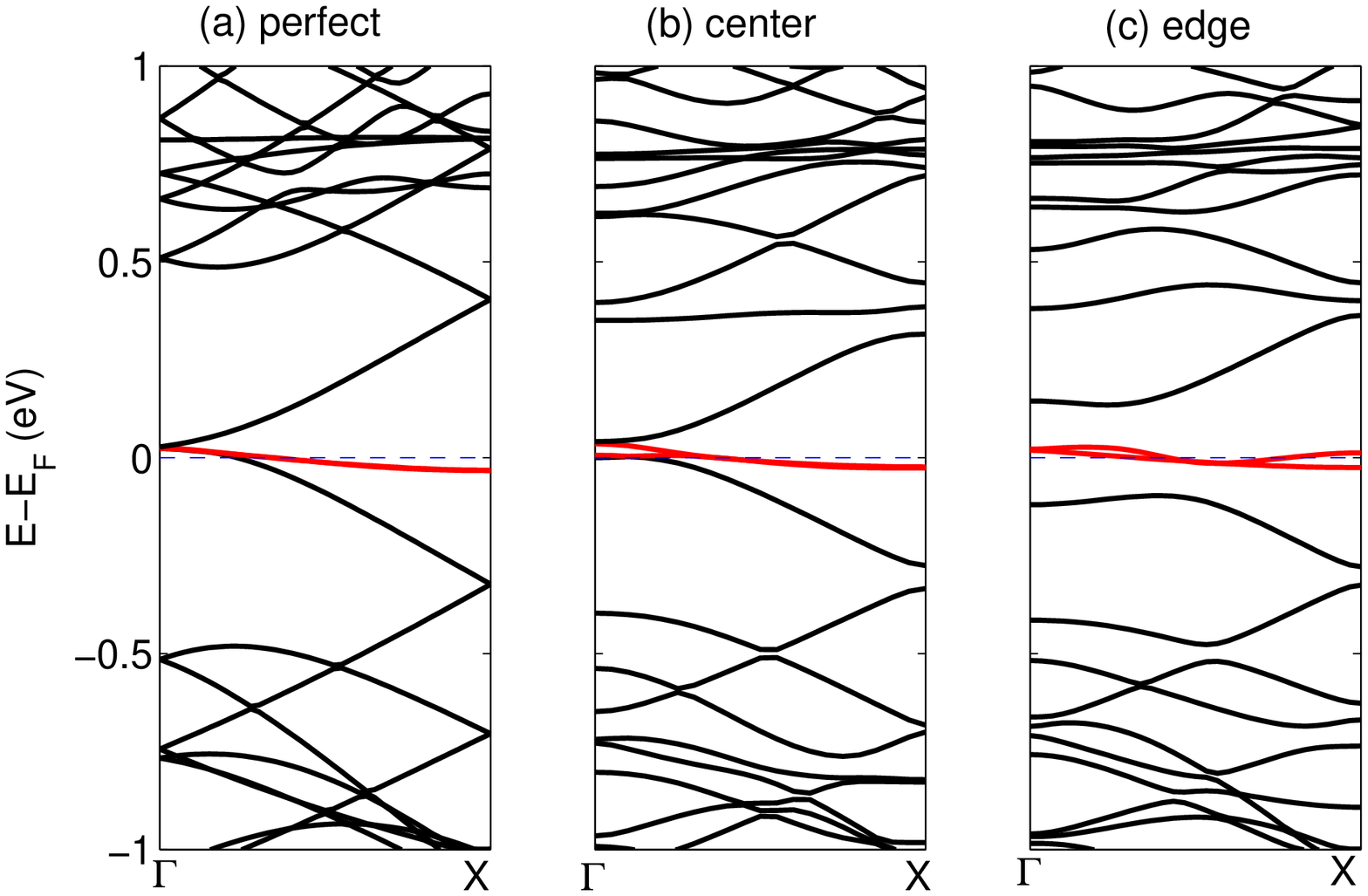}}}
\centerline{\scalebox{0.48}[0.48]{\includegraphics{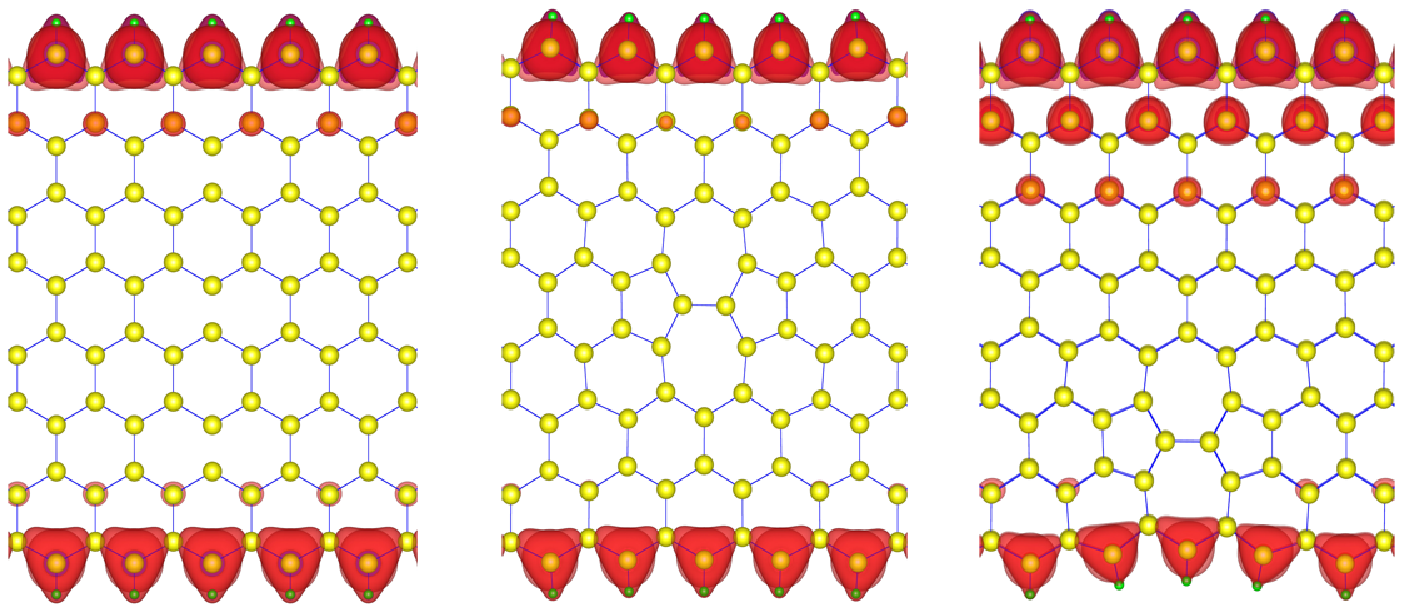}}}
\makeatother
\caption{\footnotesize{(Color online)  The band structures of ZSiNR-16 for $5\times1$ supercell of (a) perfect, (b) SW defect at the center, and (c) SW defect at the edge. The solid red lines indicate the edge state. Band decomposed charge densities of the edge states are also shown at the bottom of corresponding band structures (the isosurface is set to $0.55\times 10^{-3} $ e/$\mathrm{\AA} ^3$). }  }
\label{band16-5}

\vspace{0.1cm}
\makeatletter
\def\@captype{figure}
\centerline{
\scalebox{0.4}[0.4]{\includegraphics{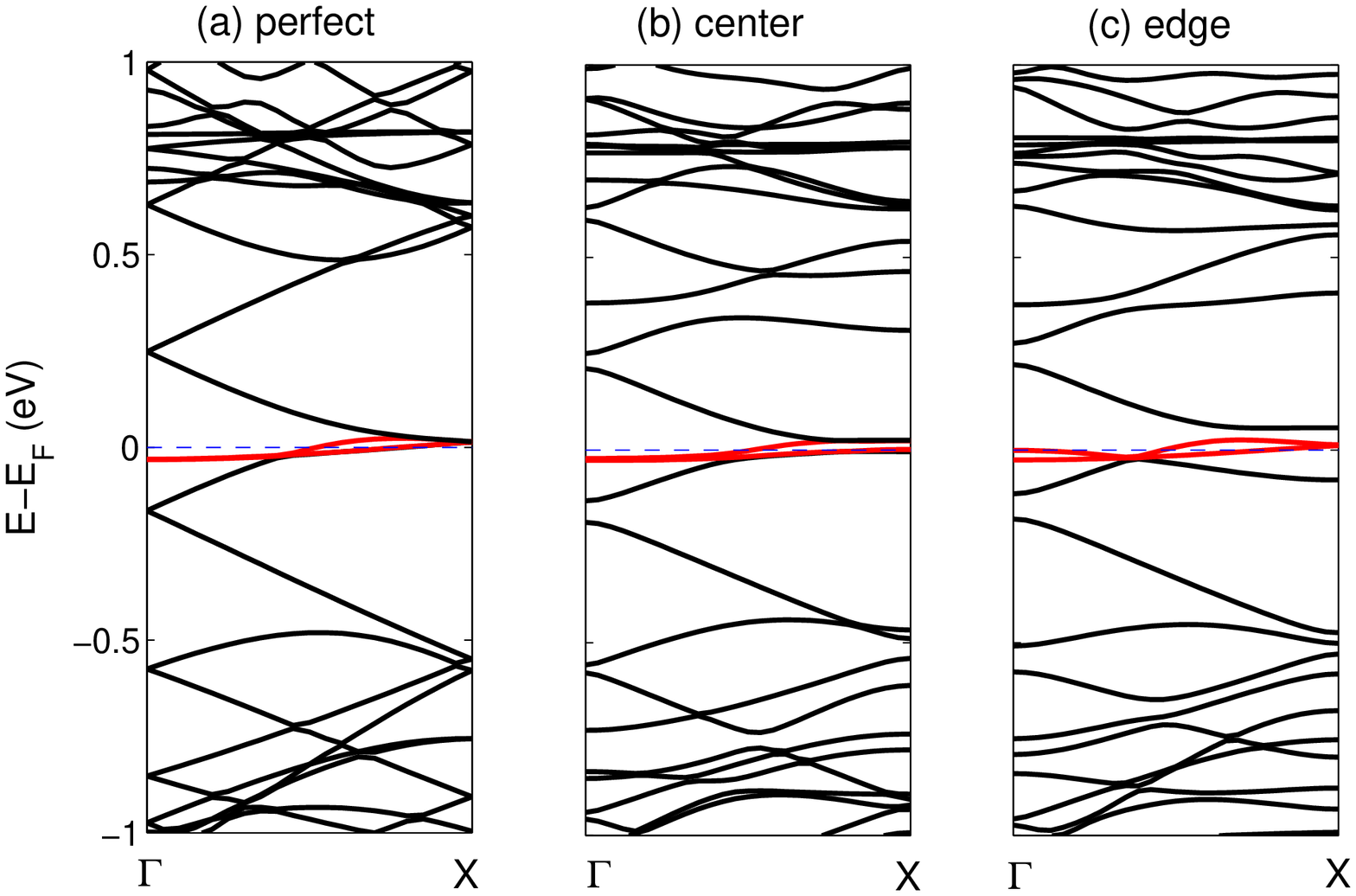}}}
\centerline{\scalebox{0.48}[0.48]{\includegraphics{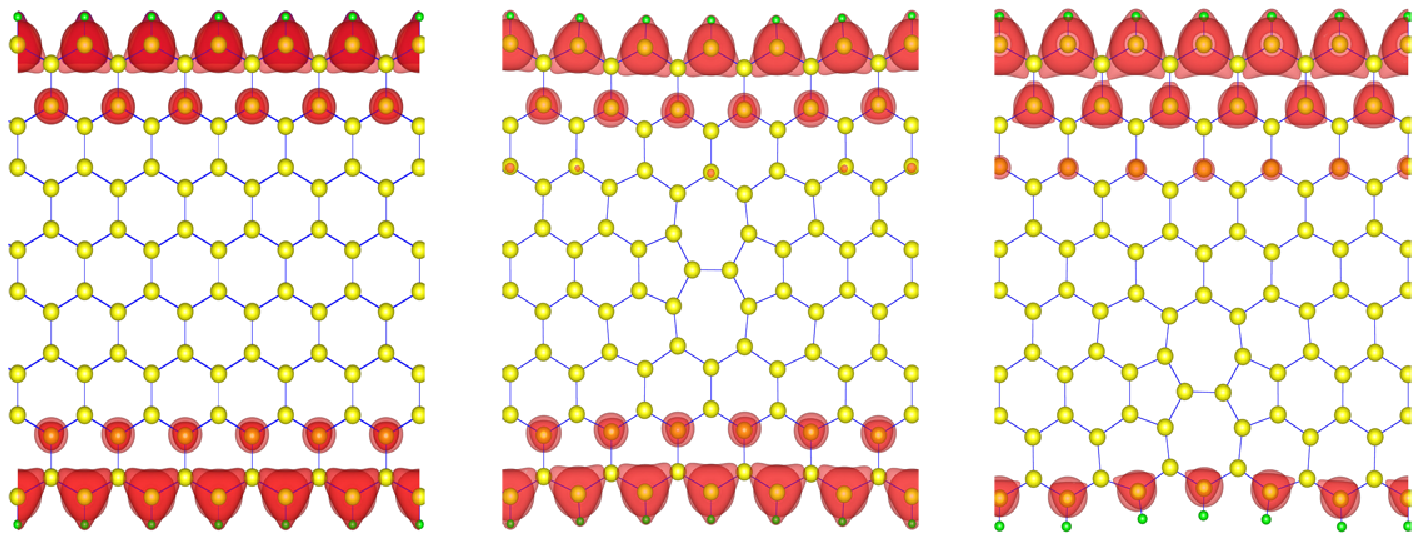}}}
\makeatother
\caption{\footnotesize{(Color online)  The band structures of ZSiNR-16 for $6\times1$ supercell of (a) perfect, (b) SW defect at the center, and (c) SW defect at the edge. The solid red lines indicate the edge state. Band decomposed charge densities of the edge states are also shown at the bottom of corresponding band structures (the isosurface is set to $0.55\times 10^{-3} $ e/$\mathrm{\AA} ^3$). }  }
\label{band16-6}

The electronic band structure of perfect ZSiNR-16, SW defect at the center, and SW defect at the edge are presented for $4\times1$, $5\times1$,  and $6\times1$ supercells are listed in FIG. \ref{band16-4}, FIG. \ref{band16-5}, and FIG. \ref{band16-6}, respectively. For each size of supercell, the band dispersion for perfect ZSiNR-16 arises from the zone folding of the NM band structure for the primitive cell of ZSiNR-16 as shown in FIG. \ref{cellribbon} (b). The flat region across the Fermi level represents edge states which are mainly localized at the edge Si atoms.  For the SW defect at the center, a band gap ~0.17 eV opens and the edge localized state occurs in $4\times1$ supercell, as well as the band structure are still metallic for $5\times1$ and $6\times1$ supercell. The band gap in in $4\times1$ supercell is due to the potential of periodic array of SW defects, but these potentials in $5\times1$ and $6\times1$ supercell are weak and thus slightly affect the electronic properties. The band structures have been strongly affected by the SW defect at the edge. The band gaps are sizable for all three sizes of supercell and decrease with increasing the supercell size, and the values are 0.33 eV, 0.23 eV , and 0.15 eV for $4\times1$, $5\times1$, and $6\times1$ supercells, respectively. In comparison with degenerate edge states of SW defect at the center, the band dispersions of edge localized states are splitting because of the symmetry broken of two edges of ZSiNR-16 with SW defect at the edge.

In order to further investigate the effects of SW defect on the electronic properties of SiNRs, the band decomposed charge densities of edge states are also shown at the bottom of corresponding band structures in FIG. \ref{band16-4}, FIG. \ref{band16-5}, and FIG. \ref{band16-6}, respectively. The results show that the SW defects at the center slightly influence on the distribution of charge density for all calculated concentrations of SW defects, in which the symmetries of two edges are remained as similar as the cases without defects.  The charge densities of the edge states at the side with SW defect transfer to the other side without defect. The splitting of degenerate edge states for the SW defect at the edge result from perturbation repulsive potential, which comes from the compressive stress field of dipole consisted of neighboring dislocation and anti-dislocation.

\vspace{0.7cm}
\makeatletter
\def\@captype{figure}
\centerline{
\scalebox{0.42}[0.42]{\includegraphics{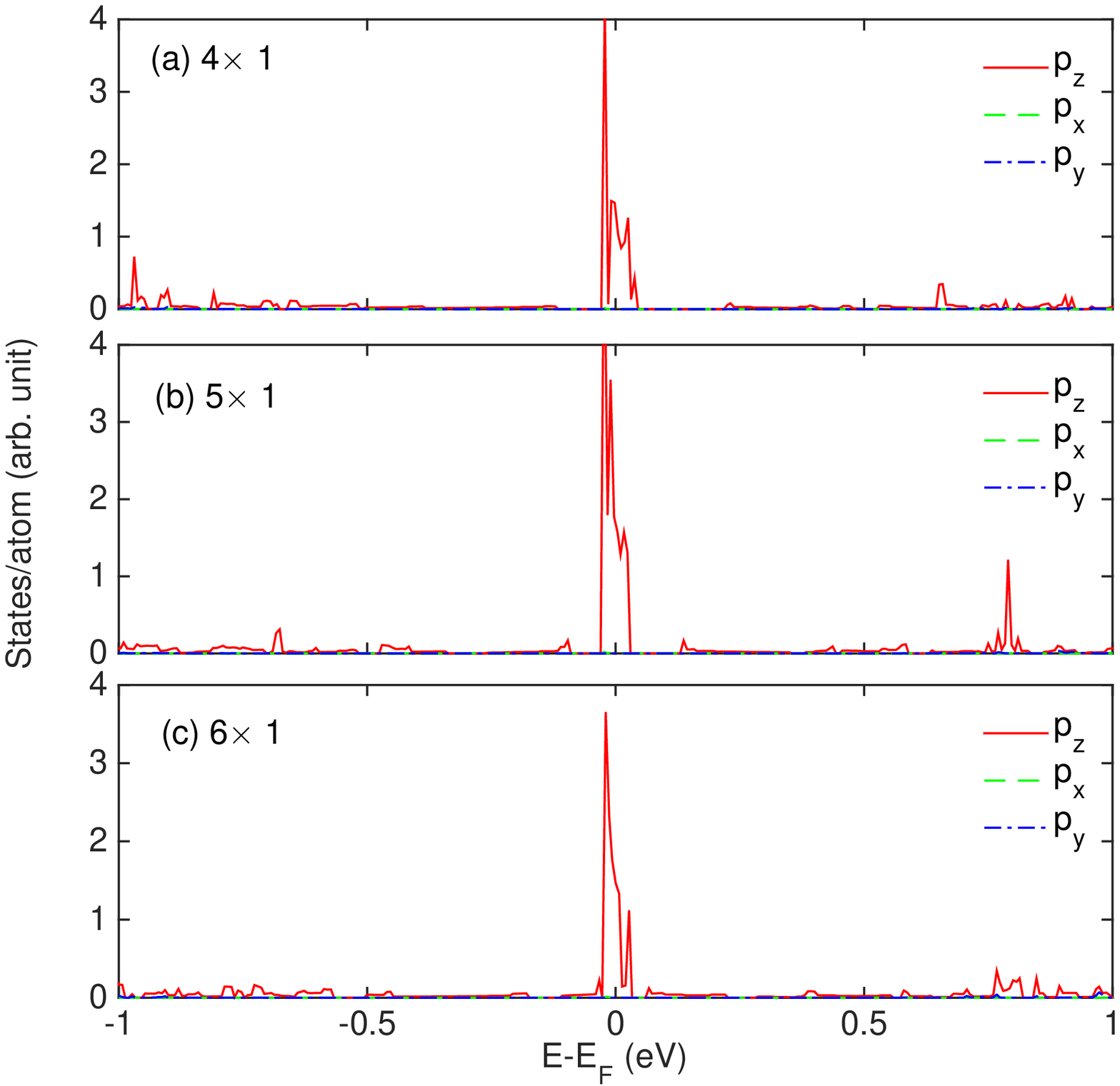}}}
\makeatother
\caption{\footnotesize{(Color online)  The partial electronic DOSs of SW defects at the edge for (a) $4\times1$, (b) $5\times1$, and (c) $6\times1$ supercells, respectively. }  }
\label{nanodos}

FIGs. \ref{nanodos}(a)-(c) list the partial electronic DOSs of SW defects at the edge for $4\times1$, $5\times1$, and $6\times1$ supercells, respectively. For all cases, the $\pi$ orbitals of the silicon atoms ($p_z$) show large peaks near the Fermi level, whereas the $p_x$ and $p_{y}$ electrons shows few contributions. In comparison with the electronic DOS of perfect ZSiNR which shows a single peak at Fermi level as shown in FIG. \ref{cellribbon}(c), the multi-peaks are present near the Fermi level in the SW-defected ZSiNRs. In addition, the locations of large peaks are always lower than Fermi level, and the results means that the electrons of conduction band mainly come from the side of ZSiNRs without SW defects.

\section{Conclusion}\label{con}
In conclusion, the structural and electronic properties of SW defects in bulk silicene and its ZSiNRs-16 with hydrogenated edges are investigated using first-principles calculations based on DFT. For all the calculations, the periodic boundary conditions are employed and a array of SW defects are considered. The formation energies and structural parameters of SW defects in bulk silicene depending on the concentration of defects are calculated both in GGA and LDA formalisms. Our results show a good agreement with available values from the previous first-principles calculations. The SW-defected electronic band dispersions of $5\times5$ and $6\times6$ supercells in bulk silicene show a tiny band gap opening, because the defect breaks the symmetry and a periodic potential results from the array of SW defects. The energetics, structural aspects, and electronic properties  of SW defects locating at the center and edge of the edge-hydrogenated ZSiNR-16 are mainly investigated, and three concentrations of defects such as SW defects in $4\times1$, $5\times1$, and $6\times1$ supercells are considered, respectively. The formation energies at the edge are always lower than those at the center so the SW defects prefer to locate at the edge of ZSiNRs. The SW defects at the center of ZSiNRs slightly influence on the electronic properties, whereas the SW defects at the edge of ZSiNRs break the degenerate edge states and induce a sizable gap (0.33 eV, 0.23 eV , and 0.15 eV for $4\times1$, $5\times1$, and $6\times1$ supercells, respectively). In addition, it is worth to find that the SW defects produce a perturbation repulsive potential, which leads the decomposed charge of edge states at the side with defect to transfer to the other side without defect. Therefore, through carefully controlling the position concentration in SiNRs, our results are potentially used to design the electronic devices based on silicene in future band engineering.

\section*{ACKNOWLEDGMENTS}
The work is supported by the National Science Foundation of China (Grant no.
11304403) and Project no. 06112015CDJXY300006 supported by the Fundamental
Research Funds for the Central Universities of China.

%\bibliography{Multiferroic_20070522}% Produces the bibliography via BibTeX.

\begin{thebibliography}{35}
\expandafter\ifx\csname
natexlab\endcsname\relax\def\natexlab#1{#1}\fi
\expandafter\ifx\csname bibnamefont\endcsname\relax
  \def\bibnamefont#1{#1}\fi
\expandafter\ifx\csname bibfnamefont\endcsname\relax
  \def\bibfnamefont#1{#1}\fi
\expandafter\ifx\csname citenamefont\endcsname\relax
  \def\citenamefont#1{#1}\fi
\expandafter\ifx\csname url\endcsname\relax
  \def\url#1{\texttt{#1}}\fi
\expandafter\ifx\csname
urlprefix\endcsname\relax\def\urlprefix{URL }\fi
\providecommand{\bibinfo}[2]{#2}
\providecommand{\eprint}[2][]{\url{#2}}

\bibitem{Cahangirov2009} S. Cahangirov, M. Topsakal, E. Akt\"{u}rk, H. \c{S}ahin, and S. Ciraci,
Phys. Rev. Lett. {\bf 102},  236804 (2009).
\bibitem{Jose2012} D. Jose, and A. Datta, J. Phys. Chem. C {\bf 116}, 24639(2012).
\bibitem{Lebegue2009} S. Lebegue and O. Eriksson, Phys. Rev. B {\bf 79}, 115409 (2009)
\bibitem{Ni2012} Z. Ni, Q. Liu, K. Tang, J. Zheng, J. Zhou, R. Qin, Z. Gao, D.
Yu, and J. Lu, Nano Lett. {\bf 12}, 113 (2012).
\bibitem{Drummond2012} N. D. Drummond, V. Z\'{o}lyomi, and V. I. Fal'ko, Phys. Rev. B
{\bf 85}, 075423 (2012).
\bibitem{Ezawa2012} M. Ezawa, New J. Phys. {\bf 14}, 033003 (2012).
\bibitem{Ezawa2013prl} M. Ezawa, Phys. Rev. Lett. {\bf 110}, 026603 (2013)
\bibitem{Vogt2012} P. Vogt, P. De Padova, C. Quaresima, J. Avila, E. Frantzeskakis,
M. C. Asensio, A. Resta, B. Ealet, and G. Le Lay, Phys. Rev.
Lett. {\bf 108}, 155501 (2012).
\bibitem{Feng2012} B. Feng, Z. Ding, S. Meng, Y. Yao, X. He, P. Cheng, L. Chen,
and K. Wu, Nano Lett. {\bf 12}, 3507 (2012).
\bibitem{Jamgotchian2012} H. Jamgotchian, Y. Colignon, N. Hamzaoui, B. Ealet, J. Y.
Hoarau, B. Aufray, and J. P. Bib¡äerian, J. Phys.: Condens.Matter
{\bf 24}, 172001 (2012).
\bibitem{Ezawa2013} M. Ezawa, Euro. Phys. J. B {\bf 86}, 139 (2013).
\bibitem{Fleurence2012} A. Fleurence, R. Friedlein, T. Ozaki, H. Kawai, Y.Wang, and Y.
Yamada-Takamura, Phys. Rev. Lett. {\bf 108}, 245501 (2012).
\bibitem{Meng2013} L. Meng, Y. Wang, L. Zhang, S. Du, R. Wu, L. Li, Y. Zhang, G. Li, H. Zhou, W. A. Hofer,  et. al., Nano Lett. {\bf 13}, 685 (2013).
\bibitem{Scalise2014} D.Chiappe, E. Scalise, E. Cinquanta, C. Grazianetti, B. v. Broek, M. Fanciulli, M. Houssa, and A. Molle, Adv. Mater. {\bf 26}, 2096 (2014).
\bibitem{Takeda1994} K. Takeda and K. Shiraishi, Phys. Rev. B {\bf 50}, 14916 (1994).
\bibitem{Ding2009} Y. Ding, J. Ni, Appl. Phys. Lett. {\bf 95}, 083115 (2009).
\bibitem{Padova2010} P. D. Padova, C. Quaresima, C. Ottaviani, P. M. Sheverdyaeva, P. Moras, C. Carbone, D. Topwal, B. Olivieri, A. Kara, H. Oughaddou, B. Aufray, and G. L. Lay, Appl. Phys. Lett. {\bf 96}, 261905 (2010).
\bibitem{Padova2011} P. D. Padova, C. Quaresima, B. Olivieri, P. Perfetti, and G. L. Lay, Appl. Phys. Lett. {\bf 98}, 081909 (2011).
\bibitem{Le2014} N. B. Le, T. D. Huan, and L. M. Woods, Phys. Rev. Appl. {\bf 1}, 054002 (2014).
\bibitem{Shakouri2015} Kh. Shakouri, H. Simchi, M. Esmaeilzadeh, H. Mazidabadi, and F. M. Peeters, Phys. Rev. B {\bf 92}, 035413 (2015).
\bibitem{Yang2014} X. F. Yang, Y. S. Liu, J. F. Feng, X. F. Wang, C. W. Zhang, and F. Chi, J. Appl. Phys. {\bf 116}, 124312 (2014).
\bibitem{Ding2014}, Y. Ding, Y. Wang, Appl. Phys. Lett. {\bf 104}, 083111 (2014).
\bibitem{Samsonidze2002} G. G. Samsonidze, G. Samsonidze, and B. I.Yakobson, Phys. Rev. Lett. {\bf 88}, 065501 (2002).
\bibitem{Ewels2002} C. P. Ewels et al., Chem. Phys. Lett. { \bf 351}, 178 (2002).
\bibitem{Ma2009} J. Ma, D. Alfe, A. Michaelides, and E. Wang, Phys. Rev. B {\bf 80}, 033407 (2009).
\bibitem{Lusk2008} M. T. Lusk and L. D. Carr, Phys. Rev. Lett. {\bf 100}, 175503 (2008).
\bibitem{Ding2007} F. Ding, K. Jiao, M. Wu, and B. I. Yakobson, Phys. Rev. Lett. {\bf 98}, 075503 (2007).
\bibitem{Wang2011} S. Wang, Y. Yao, H. Zhang, and R. Wang, Physics Letters A {\bf 375}, 4109 (2011).
\bibitem{Meyer2008} J. C. Meyer, C. Kisielowski, R. Erni, M. D. Rossell, M. F. Crommie, and A. Zettl, Nano Lett. {\bf 8}, 3582 (2008).
\bibitem{Sahin2013} H. Sahin, J. Sivek, S. Li, B. Partoens, and F. M. Peeters, Phys. Rev. B {\bf 88}, 045434 (2013).
\bibitem{Shirodkar2012} S. N. Shirodkar, U. V. Waghmare, Phys. Rev. B {\bf 86}, 165401 (2012)
\bibitem{Manjanath2013} A. Manjanath and A. K. Singh, Chem. Phys. Lett. {\bf 592}, 501 (2013).
\bibitem{Gao2013} J. Gao, J. Zhang, H. Liu, Q. Zhang, and J. Zhao, Nanoscale, {\bf 5}, 9785 (2013).
\bibitem{Dong2015} H. X. Dong, D. Q. Fang, B. H. Gong, Y. Zhang, E. Zhang, and S. L. Zhang, J. Appl. Phys. {\bf 117}, 064307 (2015)
\bibitem{Hohenberg} P. Hohenberg and W. Kohn, Phys. Rev. {\bf 136}, B864 (1964).
\bibitem{Kohn} W. Kohn, L. J. Sham, Phys. Rev. {\bf 140}, A1133 (1965).
\bibitem{Kresse2} G. Kresse and J. Furthm\"{u}ller, Phys. Rev. B, {\bf 54}, 11169 (1996).
\bibitem{Kressecom} G. Kresse and J. Furthm\"{u}ller, Comput. Mater. Sci. {\bf 6}, 15 (1996).
\bibitem{Blochl} P. E. Bl$\ddot{o}$chl, Phys. Rev. B {\bf 50}, 17953 (1994).
\bibitem{Kresse4} G. Kresse, D. Joubert, Phys. Rev. B {\bf 59}, 1758 (1999).
\bibitem{Ceperley1980} D. M. Ceperley and B. J. Alder, Phys. Rev. Lett. {\bf 45}, 566 (1980).
\bibitem{Perdew1} J. P. Perdew, K. Burke, M. Ernzerhof, Phys. Rev. Lett. {\bf77}, 3865 (1996) .
\bibitem{Perdew2} J. P. Perdew, K. Burke, M. Ernzerhof, Phys. Rev. Lett. {\bf78}, 1396 (1996).
\bibitem{Monkhorst} H. J. Monkhorst, J. D. Pack, Phys. Rev. B {\bf 13}, 5188 (1976).
\bibitem{Lee2009} G. Lee, K. Cho, Phys. Rev. B {\bf 79}, 165440 (2009).








\end{thebibliography}

\end{document}